# Programmable photon pair source

Liang Cui, [1] Jinjin Wang, [1] Jiamin Li, [1] Mingyi Ma, [1] Z. Y. Ou, [2] and Xiaoying Li,[1,†]

[1] College of Precision Instrument and Opto-Electronics Engineering, Key Laboratory of Opto-Electronics Information Technology, Ministry of Education, Tianjin University, Tianjin 300072, P. R. China

[2] Department of Physics, City University of Hong Kong, 83 Tat Chee Avenue, Kowloon, Hong Kong, P. R. China

[†] xiaoyingli@tju.edu.cn

**Abstract**

Photon pairs produced by the pulse-pumped nonlinear parametric processes have been a workhorse of quantum information science. Engineering the spectral property of the photon pairs is crucial in practical applications. In this article, we demonstrate a programmable photon pair source by exploiting a two-stage nonlinear interferometer with a phase-control device. The phase-control device introduces phase shifts by a programmable phase function that can be arbitrarily defined. With a properly designed phase function, the output spectrum of the source can be freely customized and changed without replacing any hardware component in the system. In addition to demonstrating the generation of photon pairs with factorable, positively-correlated, and negatively-correlated spectra, respectively, we show that the output of the source can be tailored into multi-channel spectrally factorable photon pairs without sacrificing efficiency. Such a source, having the ability to modify the spectrum of the photon pairs at will according to the chosen application, is a powerful tool for quantum information science.

# INTRODUCTION

Quantum correlated photon pairs are essential resources that must be freely available for implementing many of the novel functions of quantum information science (QIS). Different quantum information processing tasks require photon pairs with different spectral properties. For QIS protocols involving quantum interference between different sources, such as quantum teleportation and quantum computing, photon pairs with no frequency correlation (or, with a factorable spectrum) are desirable[1–3]. For quantum optical coherence tomography, broad bandwidth photon pairs with negative frequency correlation are more suitable[4]. While for quantum enhanced positioning, photon pairs with positive frequency correlation are preferable[5, 6]. Moreover, for practical applications of QIS, high-dimensional entangled states are desirable[7, 8]. Therefore, a photon pair source that possesses the dynamic programmability or reconfigurability of its spectral property will be a powerful tool for QIS.

A popular approach to generating photon pairs is the spontaneous parametric process in $\chi^{(2)}$- or $\chi^{(3)}$-nonlinear media[9, 10]. Because of the energy and momentum conservation, the photon pairs are highly correlated in frequency and time. When a single-frequency continuous-wave laser is used as the pump, the photon pairs have a perfect negative correlation in frequency. When an ultrashort pulse train is used as the pump, although the photon pairs are usually negatively correlated, there is the possibility to alter the spectral correlations[11–13]. With great efforts put on directly generating frequency uncorrelated photon pairs to achieve single-mode operation, significant progresses have been made in tailoring spectral property by engineering the dispersion of the nonlinear media[14–18]. However, the successes are limited to specific wavelength range, especially for the frequency uncorrelated and positively correlated cases.

Recently, engineering the spectral property of a quantum state by using the nonlinear interferometers (NLIs) has attracted a lot of attention[19–22]. In the NLIs, formed by a sequential array of nonlinear medium with gaps in between filled by linear dispersive media, photon pair generation in the parametric processes is determined by the phase matching of nonlinear media, whereas the spectral shaping is achieved independently by the linear dispersive media. The first experiment of generating the frequency uncorrelated photon pairs by quantum interferometric method is realized in a two-stage

NLI formed by two identical nonlinear fibers with a linear medium of single-mode fiber in between[21]. The results indicate that the original frequency negatively correlated joint spectral function from a single-piece nonlinear fiber can be modified to nearly uncorrelated without sacrificing the collection efficiency. However, the factorability and collection efficiency of the photon pairs are slightly deviated from the ideal case due to the overlapping between two adjacent interference fringes. It has been proposed and experimentally verified that finer spectral control can be realized if the stage number of the NLI (i.e., number of nonlinear media) is greater than two[22–26]. By properly selecting the linear dispersive media and the stage number (number of nonlinear media) of NLIs[22, 24, 26], both the central wavelengths and spectral property of photon pairs can be adjusted.

At the current stage, however, photon pairs with specific spectral properties have to be realized by specifically designed components and structure of setup. For example, for the photon pairs generated from a single-piece nonlinear medium via parametric process, the spectral property is mainly determined by the dispersion of the nonlinear medium. Even for the new aforementioned interferometric approach, the linear dispersive media need to be changed or extra stages need to be added. In any case, to alter the spectral property, one needs to change the hardware of optical components. The changing process, together with the accompanied re-aligning process, hinders the flexibility of the photon pair source.

In this paper, we demonstrate a photon pair source with programmable spectral properties for the first time. The source is based on a two-stage NLI formed by two pieces of dispersion-shifted fiber as the nonlinear waveguides and a 4f-configuration with a liquid-crystal spatial light modulator as a phase-control device in between. The phase-control device can introduce arbitrary phase shifts for different wavelengths, providing more flexibility and precision in engineering the spectra of photon pairs. By loading the properly designed phase functions on the phase-control device, we can realize photon pairs with various spectral properties, including non-correlation, positive correlation, and negative correlation. Moreover, we demonstrate a multi-channel source of photon pairs with factorable spectra, which can be used to form high dimensional entanglement and has not been experimentally realized by NLI before.

## RESULTS

**Theory**

Our programmable photon pair source is based on a two-stage NLI scheme, as shown in Fig. 1(a). The scheme is pumped by a Gaussian pulse train. The two $\chi^{(3)}$-nonlinear waveguides are identical and support single-mode propagation. Photon pairs are generated via the spontaneous four-wave mixing (SFWM) process in each nonlinear waveguide. The programmable phase-control device can introduce different phase shifts at different optical frequencies, described by a phase function $\phi(\omega)$ that can be arbitrarily defined. We assume that the phase-control device has a uniform transmission efficiency of $\eta$ for all the optical fields involved in the SFWM process. We first consider the ideal case of $\eta = 1$. The two-photon term of the output state from the scheme can be written as $|\Psi_2\rangle \propto G \int d\omega_s d\omega_i F_{NLI}(\omega_s, \omega_i)|1_s, 1_i\rangle$, where $G \propto \gamma P_p L$ is the gain parameter with $\gamma$, $P_p$, and $L$ respectively denoting the nonlinear coefficient, peak pump power, and length of each waveguide; $F_{NLI}(\omega_s, \omega_i)$ is the joint spectral function (JSF) of the photon pairs from the NLI, which describes the probability amplitude of a pair of signal and idler photons emerging at frequencies $\omega_s$ and $\omega_i$, respectively. Due to the quantum interference of the parametric processes in the two waveguides[22], $F_{NLI}(\omega_s, \omega_i)$ can be expressed as [22]:

$$F_{NLI}(\omega_s, \omega_i) = F_{SP}(\omega_s, \omega_i) \times I(\omega_s, \omega_i) \qquad (1)$$

where $F_{SP}(\omega_s, \omega_i) = \exp\left[-\frac{(\omega_s+\omega_i-2\omega_{p0})^2}{4\sigma_p^2}\right]\text{sinc}\left(\frac{\Delta k L}{2}\right)e^{i\frac{\Delta k L}{2}}$ is the JSF of photon pairs produced from a single-piece nonlinear waveguide and $I(\omega_s, \omega_i) = \cos\left(\frac{\Delta k L + \Delta \phi}{2}\right)e^{i\frac{\Delta \phi}{2}}$ is the interference function. Here, $\omega_{p0}$ and $\sigma_p$ are the central frequency and bandwidth of the Gaussian pump; $\Delta k = 2k\left(\frac{\omega_s+\omega_i}{2}\right) - k(\omega_s) - k(\omega_i) - 2\gamma P_p$ is the wave vector mismatch of the pump, signal, and idler fields in the nonlinear waveguide, and $\Delta \phi = 2\phi\left(\frac{\omega_s+\omega_i}{2}\right) - \phi(\omega_s) - \phi(\omega_i)$ is the phase difference introduced by the phase-control device.

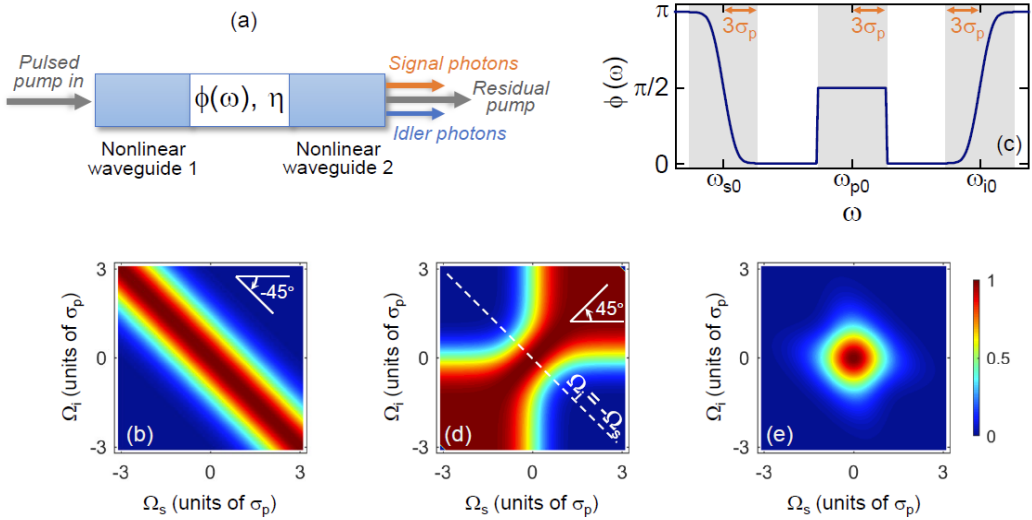

Fig. 1 (a) Two-stage nonlinear interferometer (NLI) formed by two nonlinear waveguides with a programmable phase-control device in between. (b) Contour plot of the JSF $|F_{SP}(\Omega_s, \Omega_i)|^2$ of photon pairs from a single-piece waveguide. (c) A sample phase function $\phi(\omega)$ designed for tailoring signal and idler photon pairs with factorable joint spectrum. (d) Contour plot of the interference function $|I(\Omega_s, \Omega_i)|^2 = |\cos[\frac{u(\Omega_s, a) - u(\Omega_i, a)}{2}]|^2$ with the bandwidth parameter $a = \sigma_p$. (e) Contour plot of the JSF $|F_{NLI}(\Omega_s, \Omega_i)|^2$ of the photon pairs from the NLI. $\omega_{p0}$ and $\sigma_p$ are the central frequency and bandwidth of the pump, respectively. $\omega_{s0}$ and $\omega_{i0}$ are the central frequencies of the signal and idler, respectively. In (b), (d), and (e), the substitutions $\Omega_s = \omega_s - \omega_{s0}$ and $\Omega_i = \omega_i - \omega_{i0}$ are used. The dashed line in (d) represents $\Omega_i = -\Omega_s$.

We suppose that in the waveguides the phase matching condition of SFWM is perfectly satisfied at the pump, signal, and idler frequencies, $\omega_{p0}$, $\omega_{s0}$, and $\omega_{i0}$, respectively. This means that $\Delta k = 2k(\omega_{p0}) - k(\omega_{s0}) - k(\omega_{i0}) - 2\gamma P_p = 0$ and $2\omega_{p0} = \omega_{s0} + \omega_{i0}$. Around these perfect phase matching frequencies, we have $\Delta k L \to 0$. In this case, the JSF of the single-piece nonlinear waveguide can be simplified to the Gaussian pump envelop:

$$F_{SP}(\Omega_s, \Omega_i) = \exp\left[-\frac{(\Omega_s + \Omega_i)^2}{4\sigma_p^2}\right], \tag{2}$$

and the interference function becomes

$$I(\Omega_s, \Omega_i) = \cos\left(\frac{\Delta\phi}{2}\right). \tag{3}$$

Note that we have made the substitutions: $\Omega_s = \omega_s - \omega_{s0}$ and $\Omega_i = \omega_i - \omega_{i0}$, and dropped the imaginary phase terms for simplicity. From Eq. (3), one sees that $I(\Omega_s, \Omega_i)$, the key for tailoring the final JSF $F_{NLI}(\Omega_s, \Omega_i)$, is determined only by the phase difference $\Delta\phi$ introduced by the phase-control device.

Considering the contour of $F_{SP}(\Omega_s, \Omega_i)$ has a Gaussian profile with a direction of $-45°$ from the horizontal axis (see Fig. 1(b)), we generally expect that the contour of $I(\Omega_s, \Omega_i)$ is also a Gaussian but has an orthogonal direction of $45°$, so that we can obtain $F_{NLI}(\omega_s, \omega_i)$ with various spectral properties, especially the round-shaped factorable one [13, 18]. In other words, we expect $I(\Omega_s, \Omega_i) \sim \exp[-\frac{(\Omega_s - \Omega_i)^2}{4a^2}]$ where $a$ is a bandwidth parameter describing the width of the contour. Based on this expectation, we first define $u(x, a) = \arctan(\frac{\sqrt{2}x}{a} + \frac{x^3}{\sqrt{2}a^3} + \frac{5x^5}{12\sqrt{2}a^5} + \frac{x^7}{8\sqrt{2}a^7} + \frac{79x^9}{2880\sqrt{2}a^9})$, thus we have $\cos[u(x, a)] \approx \sum_{n=0}^{4} \frac{(-1)^n x^{2n}}{n! a^{2n}} + O(x^9) \approx \exp(-x^2/a^2)$. Then we construct a piece-wise phase function $\phi(\omega)$ as shown in Fig. 1(c) to introduce different phase shifts in the pump, signal, and idler bands. The central frequencies of the pump, signal, and idler bands are $\omega_{p0}$, $\omega_{s0}$, and $\omega_{i0}$, respectively, while the half-widths of all the three bands are set to be $3\sigma_p$. Note that we should ensure that there is no overlap between different bands. For the pump band, we add a fixed phase shift of $\pi/2$; for the signal band, the introduced phase shift is described by function $\phi(\Omega_s) = -u(\Omega_s, a) + \pi/2$; for the idler band, the phase shift is $\phi(\Omega_i) = u(\Omega_i, a) + \pi/2$. Besides the three bands, we fill the undefined intervals with 0 or $\pi$ to keep the continuity of $\phi(\omega)$. From Fig. 1(c), one sees that the constructed $\phi(\omega)$ is symmetric about $\omega_{p0}$. After applying the constructed $\phi(\omega)$, the interference function becomes

$$I(\Omega_s, \Omega_i) = \cos\left[\frac{u(\Omega_s, a) - u(\Omega_i, a)}{2}\right]. \tag{4}$$

whose contour plot is shown in Fig. 1(d). One sees that the ridge of the contour is described by $u(\Omega_s, a) - u(\Omega_i, a) = 0$, solving which we get $\Omega_i = \Omega_s$. Thus the direction of the contour is $45°$.

Along the cross section line of $\Omega_i = -\Omega_s$, the interference function can be expressed as $I(\Omega_s, \Omega_i) = \cos[u(\frac{\Omega_s-\Omega_i}{2}, a)] \approx \exp[-\frac{(\Omega_s-\Omega_i)^2}{4a^2}]$, which is just what we expected.

We can control the final JSF $F_{NLI}(\omega_s, \omega_i)$ by adjusting the bandwidth parameter $a$. For example, we can obtain the round-shaped factorable JSF by setting $a = \sigma_p$ to make the contour of $I(\Omega_s, \Omega_i)$ having the same width as that of $F_{SP}(\Omega_s, \Omega_i)$ (this is exactly the case shown in Fig. 1(d)). The obtained factorable JSF is shown in Fig. 1(e). To check its factorability, we perform a singular mode decomposition[27] and find the mode number $K = 1.01$, which is very close to the ideal single-mode case. Besides the factorable JSF, we can also create a positively correlated JSF by setting $a < \sigma_p$, or, a negatively correlated one by setting $a > \sigma_p$. As examples, Figs. 2(a) and 2(d) respectively show the contours of $I(\Omega_s, \Omega_i)$ and $F_{NLI}(\omega_s, \omega_i)$ for $a = 0.5\sigma_p$, while Figs. 2(b) and 2(e) show the results for $a = 1.7\sigma_p$.

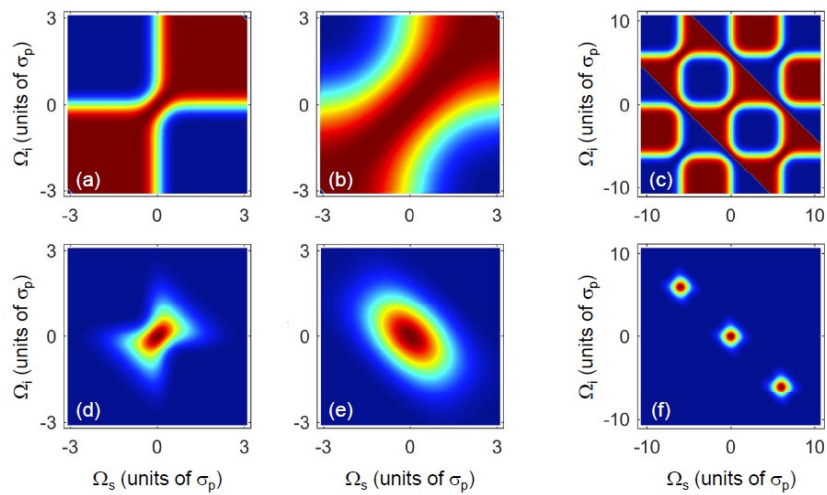

Fig. 2. Plots in the top row (a-c) are the contour plots of the interference function $|I(\Omega_s, \Omega_i)|^2 = |\cos[\frac{u(\Omega_s,a)-u(\Omega_i,a)}{2}]|^2$ and plots in the bottom row (d-f) are the contour plots of the corresponding JSF $|F_{NLI}(\Omega_s, \Omega_i)|^2$. For (a) and (d), $a = 0.5\sigma_p$; For (b) and (e), $a = 1.7\sigma_p$; (c) and (f) are the results when the phase shift patterns of the signal and idler bands in Fig. 1(c) are repeated for three times.

The above examples show that we can obtain JSFs with a customized "island" structure centering at $(\Omega_s = 0, \Omega_i = 0)$. Actually, we can create such island structure centering at any arbitrary frequencies/wavelengths within the gain bandwidth of SFWM, just like painting on a blank canvas. We can also realize JSFs with a multi-island structure by repeating the phase shift patterns in the signal and idler bands shown in Fig. 1, and the central frequencies/wavelengths and spectral correlation property of each island can be controlled independently. As an example, we create a JSF with three well-separated factorable islands, and the intensity contours of $I(\Omega_s, \Omega_i)$ and $F_{NLI}(\omega_s, \omega_i)$ are shown in Figs. 2(c) and 2(f), respectively. From Fig. 2(f), one sees that the JSF can be decomposed as $F_{NLI}(\omega_s, \omega_i) = \sum_{k=1}^{3} r_k F_{NLI}^{(k)}(\omega_s, \omega_i)$, where $\sum_{k=1}^{3} |r_k|^2 = 1$ and $F_{NLI}^{(k)}(\omega_s, \omega_i) = \psi^{(k)}(\omega_s)\phi^{(k)}(\omega_i)$ is the partial JSF for the $k$th factorable island. Since each island can be seen as an independent spectral-temporal mode, these modes are coherently superposed and form a high-dimensional entangled state

$$|\Psi\rangle \propto \sum_{k=1}^{3} \int d\omega_s d\omega_i \psi^{(k)}(\omega_s)\phi^{(k)}(\omega_i)|1_s, 1_i\rangle = \sum_{k=1}^{3} |k_s\rangle|k_i\rangle, \qquad (5)$$

where $|k_{s(i)}\rangle$ represents the $k$th mode in the signal (idler) field. Note that the state in Eq. (5) can also be generated from a cavity, and the frequency difference between different wave-packets is determined by the size of the cavity[7]. However, using our source, the frequency difference between adjacent factorable islands can be flexibility tuned by loading properly designed phase function on the phase-control device.

The above analyses have shown the effectiveness of our scheme, but there are several points that need to be elaborated further. First, in our model we have omitted the term $\Delta kL$. However, even in a case that $\Delta kL$ can not be omitted, e. g., the frequencies of the signal and idler photons are not near the perfect phase-matched frequencies, we can always compensate $\Delta kL$ by adding an opposite compensation term when constructing the phase function $\phi(\omega)$. So that the final interference function $I(\Omega_s, \Omega_i)$ can always follow our expectation. Second, in our model we have made the contour of $I(\Omega_s, \Omega_i)$ along 45° by constructing a symmetric $\phi(\omega)$. However, the phase-shift introduced in the signal and idler bands can be asymmetric as well. We can freely control the contour direction of $I(\Omega_s, \Omega_i)$ by setting different bandwidth parameters in the signal and idler bands (see Supplementary Section 1 for details). This is very usesful in tailoring the JSF when the contour of the single-piece JSF is along an arbitrary angle. Third, in practice, the ideal case of $\eta = 1$ is hard to achieve. We usually have $\eta < 1$ due to the existence of transmission loss. In the case of $\eta < 1$, the two-photon term of the output state from the scheme is evolved into (see Supplementary Section 2 for details):

$$\begin{aligned}|\Psi_2\rangle = \;&(1-\eta)G \int d\omega_s d\omega_i F_{SP}(\omega_s, \omega_i)|0_s, 0_i\rangle \\ &+\sqrt{\eta(1-\eta)}G \int d\omega_s d\omega_i F_{SP}(\omega_s, \omega_i)|1_s, 0_i\rangle \\ &+\sqrt{\eta(1-\eta)}G \int d\omega_s d\omega_i F_{SP}(\omega_s, \omega_i)|0_s, 1_i\rangle \\ &+2\eta G \int d\omega_s d\omega_i F_{NLI}(\omega_s, \omega_i)|1_s, 1_i\rangle.\end{aligned} \quad (6)$$

On the right side of Eq. (6), the first three terms arise from the transmission loss, and the last term is the two-photon state that we are interested in. The first term is the vacuum state which means both the signal and idler photons of a pair are lost due to the transmission loss, the second (third) term is the one-photon state which originates from the only surviving signal (idler) photon of a pair. The one-photon states can be seen as a background noise of the photon pairs, which can significantly influence the modal purity and collection efficiency (or heralding efficiency) of the source (see Supplement 2 for details).

Our analysis of NLI with $\chi^{(3)}$-nonliear media can be extended to other platform, such as $\chi^{(2)}$-nonlinear media[28, 29]. However, in the case of non-ideal transmission efficiency $\eta < 1$, additional attention should be paid in optimizing the parameters of NLI. The key to achieve a complete two-photon interference in NLI is that the gain parameters of waveguide 1 and 2, $G$ and $G'$, should have the relation $G' = \eta G$ (see Supplementary Section 2 for details). For $\chi^{(3)}$-nonlinear media, this relation is automatically satisfied when the lengths of the two nonlinear media are equal, because the pump power of the second fiber is also reduced by a factor of $\eta$. But for $\chi^{(2)}$-nonlinear waveguides, the length of each medium should be properly adjusted to meet the requirement of $G' = \eta G$.

**Experiment**

As shown in Fig. 3, we experimentally implement the nonlinear interferometer scheme by employing two identical 30-m-long DSFs as the nonlinear wave guide. The phase-control device in our nonlinear interferometer is realized by a 4f configuration consisting of two diffraction gratings, two cylindrical lenses, and a spatial light modulator (SLM). The SLM (Holoeye Pluto-2) is based on a reflective liquid crystal micro-display with a resolution of $1920 \times 1080$ pixels. Note that in Fig. 3 the SLM is depicted as a transmissive device for clarity. The light of different wavelengths from the output of the first DSF is spatially separated by the first grating and focused on different columns of pixels of the SLM by the first cylindrical lens. The SLM can introduces programmed phase shifts for different wavelengths when a properly designed gray-level pattern is loaded. Then the light of different wavelengths is recombined by using a second group of cylindrical lens and grating and then sent into the second DSF. The transmission loss in each DSF is negligible. The total transmission efficiency of the phase-control device between the two DSFs is 60%.

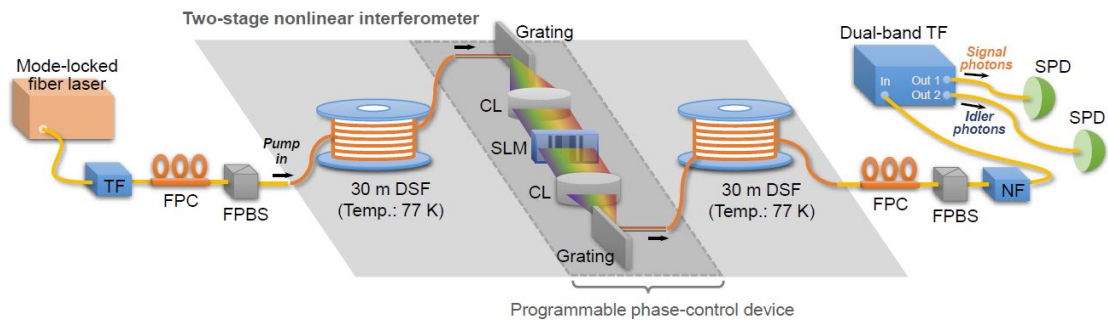

Fig. 3. Experimental setup. DSF, dispersion-shifted fiber; SLM, spatial light modulator; CL, cylindrical lens; TF, tunable filter; FPC, fiber polarization controller; FPBS, fiber polarization beam splitter; NF, notch filter; SPD, single photon detector.

We employ a telecom-band mode-locked fiber laser with a repetition rate of 36.9 MHz as the pump source, and a tunable filter (TF) realized by diffraction gratings to control the central wavelength and bandwidth of the pump. The fiber polarization controller (FPC) and polarization beam splitter (FPBS) are used for polarization purification and power control. The photon pairs generated via SFWM in the DSFs are co-polarized with the pump[30]. Therefore, we use another set of FPC and FPBS at the output of the nonlinear interferometer to select the SFWM photons and suppress the photons from spontaneous Raman scattering (SRS) that are cross-polarized with the pump[31]. Then we use a notch filter (NF) to reject the residual pump. To characterize the spectral profile of the photon pairs, the signal and idler photons are separated and selected by a dual-band TF. Two superconducting nanowire single-photon detectors (SPDs) are utilized to detect the signal and idler photons, respectively. The total detection efficiencies (including the filters and SPDs) for the signal and idler photons are both ~15%. We use a computer-controlled data acquisition system to process the detection signals. The single-channel counting rates and two-fold coincidence counting rates for photons originated from the same pulse and adjacent pulses are recorded.

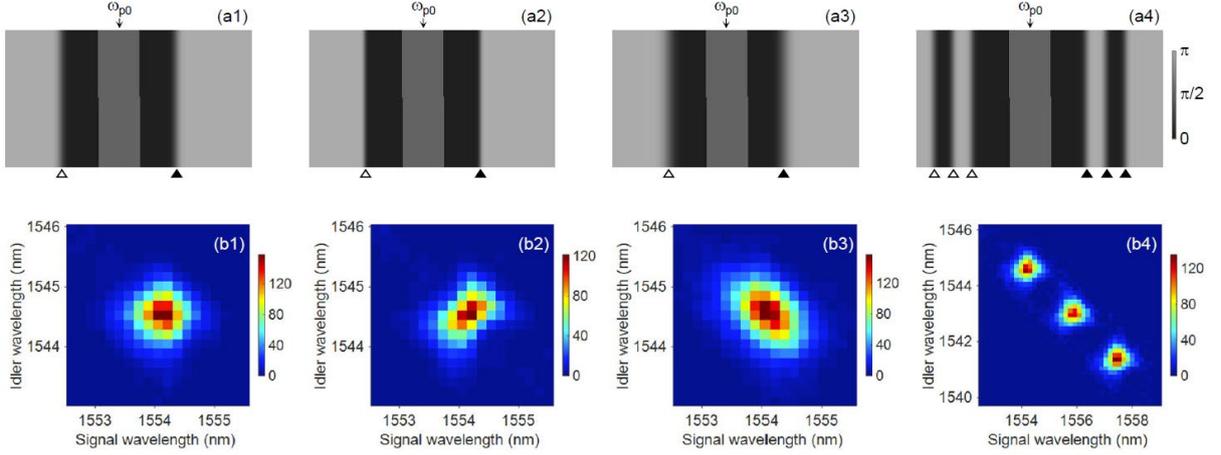

Fig. 4. (a1-a4) Gray-level patterns constructed for the SLM. (b1-b4) Measured joint spectra of photon pairs when the patterns in (a1) to (a4) are respectively loaded on the SLM. In (a1-a4), the arrows mark the columns corresponding to the central frequency/wavelength of the pump, and the solid (hollow) triangles mark the columns corresponding to the central frequencies/wavelengths of the signal (idler) photons.

In the first place, we demonstrate the generation of photon pairs with different kinds of spectral properties. The DSFs are immersed in liquid nitrogen (temperature: 77 K) to suppress SRS[30]. The zero-dispersion wavelength of the DSFs is ~1548.5 nm, so we set the central wavelength of the pump to be 1549.32 nm (193.5 THz in frequency) with a full width at half maximum (FWHM) of 0.9 nm. In this case, the phase matching condition $\Delta kL \to 0$ is nearly satisfied when the signal/idler wavelength is around the pump wavelength within tens of nanometers. Without loss of generality, we choose the central wavelengths (frequencies) of the signal and idler photons to be 1554.13 nm (192.9 THz) and 1544.53 nm (194.1 THz), respectively. We first generate photon pairs with a factorable JSF. Following the analysis represented by Fig. 1(c), we create a phase function with parameter $a = \sigma_p = 0.042$ THz, and construct the corresponding gray-level pattern as shown in Fig. 4(a1). We load the gray-level pattern on the SLM and measure the JSF of the signal and idler photons (see the joint spectral measurement in "Methods"). From the results shown in Fig. 4(b1), one sees that the measured JSF has a round shape, which agrees well with the theoretical predicted result in Fig. 1(e). Then we change the bandwidth parameter $a$ to 0.21 THz and 0.71 THz, respectively, and construct the two gray-level patterns shown in Figs. 4(a2) and 4(a3). For each case, we load the pattern on the SLM and repeat the joint spectrum measurement. The measured JSFs are shown in Figs. 4(b2) and 4(b3). As expected, their spectra exhibit positive and negative correlations, respectively.

We then demonstrate the multi-channel output feature of our source. As an example, we generate photon pairs with a JSF consisting of three separated factorable islands. The central wavelengths of the islands are specified to match the standard grid of the wavelength division multiplexing in fiber optical communication system. For the signal (idler) band, the central wavelengths of the three islands are 1554.13 (1544.53), 1555.75 (1542.94), and 1557.36 (1541.35) nm, respectively, while the corresponding frequencies are 192.9 (194.1), 192.7 (194.3), and 192.5 (194.5) THz, respectively. To achieve this, we construct the gray-level pattern shown in Fig. 4(a4). One sees that the patterns of the signal and idler bands in Fig. 4(a1) are repeated for three times to create the three islands. Compared with that for the other islands, the patterns for the central island are reversed for continuity. The measured JSF after loading the pattern on the SLM is shown in Fig. 4(b4). One sees that the three round-shaped islands are perfectly sitting at the designed wavelengths/frequencies.

Finally, we characterize the modal purity of the round-shaped factorable JSF shown in Fig. 4(b1) by measuring the second-order correlation function $g^{(2)}$ of the individual signal (or idler) field[22]. The individual signal or idler field generated by SFWM is in thermal state, its mode number $K$ is related to $g^{(2)}$ through $g^{(2)} = 1 + 1/K$[32]. Using the equations derived in Supplementary Section 2, we calculate $g^{(2)}$ as a function of the transmission efficiency $\eta$. From the results shown by the curve in Fig. 5, one sees that the value of $g^{(2)}$ can reach 1.99 in the ideal case of $\eta = 100\%$, but decreases significantly with the decrease of $\eta$. This result shows that $\eta$ has a crucial influence on the performance of our scheme. In the experiment, the round-shaped island is carved out by setting the bandwidths of both channels of the dual-band TF to 1.6 nm (200 GHz in frequency). We then conduct the $g^{(2)}$ measurement for the signal field. From the results shown by the solid circles in Fig 5, one sees that the measured $g^{(2)}$ has the same trend as the theoretical curve, but is always lower than the curve. We think that the main factor accounted for this deviation is the existence of the SRS photons, which have a different modal structure from the SFWM photons[33, 34], and another factor is the imperfection of the components (such as the limited pixel resolution of the SLM). However, the photons from SRS can be almost entirely eliminated by further cooling the fiber[35], while the performance of the components can also be improved further. We believe that the measured $g^{(2)}$ could approach the theoretical prediction after addressing the two factors. The only crucial issue is the transmission efficiency $\eta$. Currently, the efficiency $\eta$ is limited by the efficiencies of the components in the phase-control device used in our experiment. It is possible to achieve $\eta > 85\%$ by using high efficiency components (e. g., transmission grating with efficiency $> 94\%$) and optimizing the configuration. From the theoretical curve in Fig. 5, one sees that we can get $g^{(2)} > 1.89$ when $\eta > 85\%$.

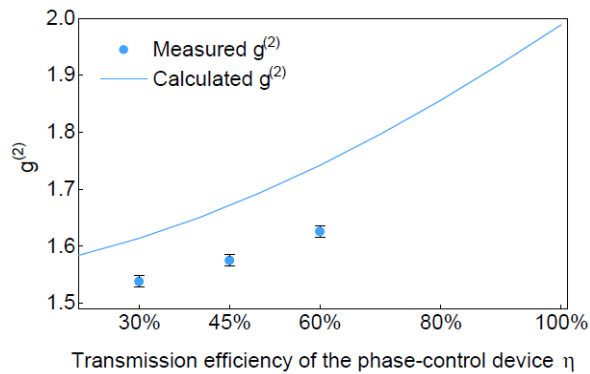

Fig. 5. Second-order correlation function $g^{(2)}$ of the individual signal field versus the transmission efficiency of the phase-control device $\eta$. The round points are the measured results of the round-shaped factorable JSF, while the solid curve is the corresponding theoretical prediction.

**CONCLUSION**

To conclude, we demonstrate a spectrally programmable photon pair source by using a pulse pumped two-stage NLI scheme, in which two pieces of DSF are utilized as the nonlinear media for SFWM and a 4f-configuration with a liquid crystal micro-display SLM is used as the programmable phase-control device. We can customize the spectral properties of the photon pairs by loading a properly designed phase function on the phase-control device. To change the output property of the source, we only need to re-design the loaded phase function, without replacing any components in the scheme. Moreover, the source is compatible with fiber communication wavelength division multiplexing, since it is straight forward to alter the central wavelength and bandwidth of each output channel. The changing can be very fast since the typical frame rate of the micro-display SLM can be greater than 60 Hz. Many novel functions can be exploited using our source, such as multi-channel output that can be used to explore high-dimensional entanglements[7]. Our investigation also shows that the transmission loss between the two stages of the NLI is a crucial factor influencing the performance of the source. We believe that our scheme can be further transplanted to other platforms of quantum light generation, and provide a flexible, precise, and reliable tool for quantum information processing.

## METHOD

**Joint spectral measurement for the photon pairs**

In the joint spectral measurement for the photon pairs, we set the transmission profiles of both the signal and idler channels of the dual-band TF (realized by using Finasar Waveshaper 4000) to be flattop with a fixed full-width of 0.2 nm. Then we scan the central wavelength of the signal (idler) channel from 1552.6 nm to 1555.6 nm (1543.0 nm to 1546.0 nm) by a step of 0.2 nm. At each step, we record the coincidence rate between the signal and idler photons. During this process, the average pump power is fixed at 0.7 mW. After subtracting the accidental coincidence from the signal and idler photons originated from adjacent pump pulses, we obtain the true coincidence rate per second as function of the signal and idler wavelengths. Then we can plot the two-dimensional contours shown in Fig. 4, which reflect the intensity distribution of the JSF[36].

**Second order correlation function measurement for the individual signal field**

We measure the second order correlation function $g^{(2)}$ for the individual signal photons by using the

HBT interferometer setup (not shown in Fig. 3). In the measurement, the signal photons are sent into a 50/50 fiber coupler whose two output ports are fed into two SPDs. We record the coincidence count rates between the two SPDs, including the coincidence produced by signal photons from the same pump pulses as well as the accidental coincidence produced by signal photons from adjacent pump pulses. Then we can obtain $g^{(2)}$ by calculating the ratio between the measured coincidence and accidental coincidence count rates.

**DATA AVAILABILITY**

The data that support the findings of this study are available from the corresponding author upon reasonable request.

## ACKNOWLEDGEMENTS

This work was supported by the Science and Technology Program of Tianjin (18ZXZNGX00210) and the National Natural Science Foundation of China (11874279, 12074283, 11527808).

## AUTHOR CONTRIBUTIONS

The work was conceived by X. L., L. C. and Z. Y. O.. L. C., J. W. and M. M. conducted the experimental work. J. L., L. C. and Z. Y. O. conducted the theoretical work. X. L. supervised this project. All authors wrote and edited the manuscript.

## CONFLICT OF INTEREST

The authors declare that they have no conflict of interest.